\documentstyle[12pt,epsf,rotating]{article}
\catcode`@=11
\def\citer{\@ifnextchar [{\@tempswatrue\@citexr}{\@tempswafalse\@citexr[]}}

\def\@citexr[#1]#2{\if@filesw\immediate\write\@auxout
        {\string\citation{#2}}\fi
\def\@citea{}\@cite{\@for\@citeb:=#2\do
        {\@citea\def\@citea{--}\@ifundefined
        {b@\@citeb}{{\bf ?}\@warning
        {Citation `\@citeb' on page \thepage \space undefined}}
        {\csname b@\@citeb\endcsname}}}{#1}}
\catcode`@=12

\def\fo{\hbox{{1}\kern-.25em\hbox{l}}}
\def\fnote#1#2{\begingroup\def\thefootnote{#1}\footnote{#2}\addtocounter
{footnote}{-1}\endgroup}

\renewcommand{\thefootnote}{\fnsymbol{footnote}}

\def\beq{\begin{equation}}
\def\eeq{\end{equation}}
\def\eq{\end{equation}}
\def\to{\rightarrow}

\def\bsg{\ifmmode B\to X_s\gamma\else $B\to X_s\gamma$\fi}
\def\bsll{\ifmmode B\to X_s\ell^+\ell^-\else $B\to X_s\ell^+\ell^-$\fi}
\def\bstt{\ifmmode B\to X_s\tau^+\tau^-\else $B\to X_s\tau^+\tau^-$\fi}
\def\shat{\ifmmode \hat{s}\else $\hat{s}$\fi}

\newcommand{\newc}{\newcommand}

\newc{\lcal}{\int {\cal L}dt}

\newc{\mHpm}{m_{H^\pm}}
\newc{\gsim}{\lower.7ex\hbox{$\;\stackrel{\textstyle>}{\sim}\;$}}
\newc{\lsim}{\lower.7ex\hbox{$\;\stackrel{\textstyle<}{\sim}\;$}}
\newc{\ie}{{\it i.e.}}          
\newc{\etal}{{\it et al.}}
\newc{\eg}{{\it e.g.}}          
\newc{\kev}{\hbox{\rm\,keV}}            
\newc{\mev}{\hbox{\rm\,MeV}}            
\newc{\gev}{\hbox{\rm\,GeV}}            
\newc{\tev}{\hbox{\rm\,TeV}}
\newc{\xpb}{\hbox{\rm\, pb}}
\newc{\xfb}{\hbox{\rm\, fb}}

\newc{\mtop}{m_t}
\newc{\mbot}{m_b}
\newc{\mz}{m_Z}
\newc{\mw}{M_W}
\newc{\alphasmz}{\alpha_s(m_Z^2)}
\newc{\swsq}{\sin^2\theta_W}
\newc{\tw}{\tan\theta_W}
\newc{\cw}{\cos\theta_W}
\newc{\sw}{\sin\theta_W}
\newc{\BR}{\hbox{\rm BR}}
\newc{\zbb}{Z\to b\bar}
\newc{\Gb}{\Gamma (Z\to b\bar b)}
\newc{\Gh}{\Gamma (Z\to \hbox{\rm hadrons})}
\newc{\rbsm}{R_b^\hbox{\rm sm}}
\newc{\rbsusy}{R_b^\hbox{\rm susy}}
\newc{\drb}{\delta R_b}

\newc{\sgn}{\mbox{sgn}}
\newc{\tbeta}{\tan\beta}
\newc{\uL}{{\tilde u_L}}
\newc{\uR}{{\tilde u_R}}
\newc{\cL}{{\tilde c_L}}
\newc{\cR}{{\tilde c_R}}
\newc{\tL}{{\tilde t_L}}
\newc{\tR}{{\tilde t_R}}
\newc{\dL}{{\tilde d_L}}
\newc{\dR}{{\tilde d_R}}
\newc{\sL}{{\tilde s_L}}
\newc{\sR}{{\tilde s_R}}
\newc{\bL}{{\tilde b_L}}
\newc{\bR}{{\tilde b_R}}
\newc{\eL}{{\tilde e_L}}
\newc{\eR}{{\tilde e_R}}
\newc{\mhp}{m_{H^\pm}}
\newc{\mhalf}{m_{1/2}}

\newc{\lR}{\tilde{l}_R}
\newc{\lL}{\tilde{l}_L}
\newc{\nL}{\tilde{\nu}_L}
\newc{\na}{\chi^0_1}
\newc{\nb}{\chi^0_2}
\newc{\nc}{\chi^0_3}
\newc{\nd}{\chi^0_4}
\newc{\ca}{\chi^{\pm}_1}
\newc{\cb}{\chi^{\pm}_2}
\newc{\camp}{\chi^\mp_1}
\newc{\cbmp}{\chi^\mp_1}
\newc{\capos}{\chi^{+}_1}
\newc{\caneg}{\chi^{-}_1}
\newc{\phit}{\phi_t}
\newc{\phib}{\phi_b}
\newc{\phiew}{\phi_{ew}}
\newc{\htz}{h^0_t}
\newc{\hbz}{h^0_b}
\newc{\hewz}{h^0_{ew}}
\newc{\hsmz}{h^0_{SM}}

\def\NPB#1#2#3{Nucl. Phys. B {\bf #1} (19#2) #3}
\def\PLB#1#2#3{Phys. Lett. B {\bf #1} (19#2) #3}

\def\PRD#1#2#3{Phys. Rev. D {\bf #1} (19#2) #3}

\def\ZPC#1#2#3{Zeit. f\"ur Physik C {\bf #1} (19#2) #3}

\def\beq{\begin{equation}}
\def\eeq{\end{equation}}
\def\bea{\begin{eqnarray*}}
\def\eea{\end{eqnarray*}}
\def\slashchar#1{\setbox0=\hbox{$#1$}           % set a box for #1
   \dimen0=\wd0                                 % and get its size 
   \setbox1=\hbox{/} \dimen1=\wd1               % get size of /
   \ifdim\dimen0>\dimen1                        % #1 is bigger 
      \rlap{\hbox to \dimen0{\hfil/\hfil}}      % so center / in box 
      #1                                        % and print #1
   \else                                        % / is bigger 
      \rlap{\hbox to \dimen1{\hfil$#1$\hfil}}   % so center #1
      /                                         % and print /
   \fi}                                         %
\catcode`@=11
\long\def\@caption#1[#2]#3{\par\addcontentsline{\csname
  ext@#1\endcsname}{#1}{\protect\numberline{\csname
  the#1\endcsname}{\ignorespaces #2}}\begingroup
    \small
    \@parboxrestore
    \@makecaption{\csname fnum@#1\endcsname}{\ignorespaces #3}\par
  \endgroup}
\catcode`@=12

\def\jfig#1#2#3{
 \begin{figure}
 \centering
 \epsfysize=3.0in
 \hspace*{0in}
 \epsffile{#2}
 \caption{#3}
 \label{#1}
 \end{figure}}
\def\rfig#1#2#3{
 \begin{figure}
 \centering
 \epsfysize=4.5in
 \hspace*{-0.5in}
\begin{turn}{-90}%
 \epsffile{#2}
\end{turn}
 \caption{#3}
 \label{#1}
 \end{figure}}
\begin{document}

\begin{titlepage}

\begin{flushright}
CERN-TH/97-331 \\
SLAC-PUB-7703 \\
hep-ph/9711410\\
November 1997
\end{flushright}

%\vspace{.3in}
%\begin{center}
%ROUGH DRAFT, DO NOT DISTRIBUTE
%\end{center}

\Large
%\vspace{0.05in}
%\renewcommand{\thefootnote}{\fnsymbol{footnote}}
\begin{center}\
{\sc Higgs Bosons Strongly Coupled \\
to the Top Quark}
\end{center}
%\vspace{0.2in}

\large

\vspace{.1in}
\begin{center}

Michael Spira${}^a$ and James D. Wells${}^b$ \\

\normalsize

\vspace{.1in}
{\it 
${}^a$CERN, Theory Division, CH-1211 Geneva, Switzerland \\ 
\vspace{.1in}
${}^b$Stanford Linear Accelerator Center \\
Stanford University, Stanford, CA 94309\fnote{\dagger}{Work 
supported by the Department of Energy
under contract DE-AC03-76SF00515.}\\}

\end{center}
 
%\vspace{0.2in}
 
\vspace{0.15in}
 
\normalsize

\begin{abstract}
 
Several extensions of the
Standard Model require the burden of electroweak symmetry breaking
to be shared by multiple states or sectors.  This leads to the
possibility of the top quark interacting with a scalar more strongly
than it does with the Standard Model Higgs boson.  
In top-quark condensation this possibility is natural.
We also discuss how this might be realized in supersymmetric theories.
The properties of a strongly coupled Higgs boson in top-quark
condensation and supersymmetry are described.  We comment on the difficulties
of seeing such a state at the Tevatron and LEPII, and study the
dramatic signatures it could produce at the LHC.  The four top quark
signature is especially useful in the search for a strongly coupled
Higgs boson.  We also calculate the rates of the more conventional
Higgs boson signatures at the LHC, including the two photon and four
lepton signals, and compare them to expectations in the Standard Model.

\end{abstract}

\vspace*{\fill}

\begin{flushleft}
CERN-TH/97-331 \\
SLAC-PUB-7703 \\
hep-ph/9711410\\
November 1997
\end{flushleft}

\end{titlepage}

\baselineskip=18pt
%\baselineskip=33pt

%\setcounter{footnote}{0}
%\setcounter{page}{1}
%\setcounter{figure}{0}
%\setcounter{table}{0}

%%%%%%%%%%%%%%%%%%%%%%%%%%%%%%%%%%%%%%%%%%%%%%%%%%%%%%%%%%%%%%%%%%%%%%%
%sss

\section{Introduction}

Electroweak symmetry breaking and fermion mass generation are both
not understood.  The Standard Model (SM) with one Higgs scalar doublet
is the simplest mechanism one can envision.  The strongest arguments
in support of the Standard Model Higgs mechanism is that no experiment
presently refutes it, and that it allows both fermion masses and electroweak
symmetry breaking. 
Perhaps the most crucial test of this standard postulate will be
its confrontation with a large number of top quark events.
The top quark, being the heaviest known chiral fermion, 
may be the most sensitive to dynamics
which produce fermion masses but have little to do with electroweak
symmetry breaking.  

One consequence of the existence of a Higgs boson strongly coupled
to the top quark is that a perturbative description up
to the Planck scale is probably not possible.  If the top quark
Yukawa coupling to the Higgs boson exceeds about $1.3$ at the weak scale
it will
diverge before reaching the Planck scale.  The specific scale
at which a Landau pole develops for a particular value of the
Yukawa coupling depends on the gauge symmetries and particle content
of the underlying low energy theory. For top quark
condensate models~\citer{r2,r49}, 
this Landau pole development at low scales
is welcomed in order to satisfy constituent
relations between the low scale effective theory and the more
fundamental theory (e.g., topcolor)~\citer{r53,r13}. 
Strong coupling dynamics
are crucial for the success of these theories.  For supersymmetry,
non-perturbative dynamics are often frowned upon by model builders
since one generally loses control over the successful prediction
of $\sin^2\theta_W$~\cite{r33}, which seems to require perturbative evolution
from the grand unified scale down to the weak scale.  However,
there are numerous examples now of strongly coupled supersymmetric
theories that do not disrupt gauge
coupling unification~\cite{r34}.  We therefore do not consider gauge coupling
unification to be a necessary impediment to strongly coupled
Higgs bosons.

Furthermore, it is tempting to consider the top quark as the only
fermion with an understandable mass since it has a sizeable Yukawa coupling
in the Standard Model
while the other fermions have small ones.
This attitude leads one to concentrate on finding ways
to suppress the other fermion masses, rather than to explain the
top mass.  Since we know so little about fermion mass generation
and electroweak symmetry breaking, it is perhaps dangerous to
be aesthetically anchored to this viewpoint.  A more general
approach would be to consider why the top mass is so different
than the other fermion masses.
Top-quark condensation models attempt
to answer this question, and they lead to the conclusion that the
top quark couples even more strongly to a Higgs
boson (top quark boundstate) than in the Standard Model 
\cite{r10,r13,r15}.

The prediction that the top quark
interacts more strongly in more elaborate theories than it
does in the Standard Model should not
be too surprising.
Many extensions of the Standard Model, which
can be parametrized in terms of multiple condensing fundamental scalars,
allow the top quark to interact more strongly with at least one of
these scalars than it does with the Standard Model scalar. 
This is true for top-quark condensation, and it is also frequently
true in the minimal supersymmetric Standard Model (MSSM). 

In the next two sections we will briefly discuss the properties
of a strongly interacting top Higgs particle in supersymmetric models
and top-quark condensate models.  Since top-quark condensation
more naturally yields this possibility, we will focus more intently
on it as in our example in the following section which discusses
the signals of this Higgs boson at the Large Hadron Collider (LHC).

%%%%%%%%%%%%%%%%%%%%%%%%%%%%%%%%%%%%%%%%%%%%%%%%%%%%%%%%%%%%%%%
\section{Supersymmetry}

We are interested in separating
top quark mass generation from electroweak symmetry breaking.
Although top-quark condensation will be our main illustration
for the strongly coupled Higgs boson, weak-scale supersymmetric theories
could also accomplish this and predict a strongly coupled
top Higgs boson. As we mentioned in the introduction, some of the new
developments in supersymmetric model building demonstrate how strongly
coupled theories can be desirable, and also how they can still preserve
gauge coupling unification.  Supersymmetric theories which dynamically
generate flavor from strong coupling dynamics can produce a strong top quark
Yukawa coupling at the composite scale~\cite{r51}.  
By lowering the composite scale
the top-quark Yukawa coupling at the weak scale is allowed to be much
higher than in conventional supersymmetric models that 
remain weakly coupled up to the GUT scale.  There are challenges to
constructing a model of flavor based on strong coupling dynamics.
However, the progress
made in controlling the predictions for composite models~\cite{r51}, 
along with the
realization that perturbative gauge coupling unification could still
be possible~\cite{r34} encourages us to consider a strongly coupled
Higgs boson in supersymmetry.

In the two Higgs doublet
model of the MSSM~\cite{r36} one doublet ($H_u$) gives mass to the up-type
fermions and the other ($H_d$) to the down-type fermions. 
In order for the lightest Higgs boson to couple strongly to the top-quark
the eigenvalues should be arranged such that 
$\langle H_u\rangle\ll \langle H_d\rangle$ and
$h_u^0$ is a mass eigenstate.  This arrangement, corresponding
to low $\tan\beta$, is possible as can be seen from the Higgs mass matrix in
the $\{H_d^0, H^0_u\}$ basis,
\beq
M^2= \left( \begin{array}{cc}
  m_A^2\sin^2\beta+m^2_Z\cos^2\beta & -\sin\beta\cos\beta (m_A^2+m_Z^2) \\
 -\sin\beta\cos\beta (m^2_A+m_Z^2) & m_A^2\cos^2\beta +m^2_Z\sin^2\beta
 \end{array} \right) +
 \left( \begin{array}{cc}
  \Delta_{11} & \Delta_{12} \\
  \Delta_{12} & \Delta_{22} 
 \end{array} \right) ,
\eeq
where the $\Delta_{ij}$ represent radiative corrections to the
Higgs mass matrix~\cite{r35}.
Since the tree level contribution to
$M_{12}$ becomes smaller as $\sin\beta\to 0$ and $\Delta_{12}$
generally becomes larger, cancellations between the two are possible
and can lead to a pure (or almost pure) 
$h^0_u$ mass eigenstate that is strongly
coupled to the top quark\fnote{\dagger}{Similar behavior can occur in the
$\cos\beta\to 0$ limit where $h_d^0$ becomes a heavy mass eigenstate
which couples strongly to the bottom quarks~\cite{r47a}.}.  
Furthermore, the mass of $h^0_u$ is controlled
mostly by $m_A$ (supersymmetry breaking mass scale) and $h^0_d$ by
$m_Z$ (the electroweak symmetry breaking mass scale).
The neutral Goldstone boson for general $\tan\beta$,
\beq
G^0\propto \cos\beta\ {\rm Im} (H_d^0)-\sin\beta\ {\rm Im}(H^0_u)
\eeq
approaches $G^0\sim {\rm Im}(H_d^0)$ in the $\sin\beta\to 0$ limit.

The value of $m_A$ is probably larger than $m_Z$ given current
experimental limits on other sparticle masses dependent on 
the overall scale of supersymmetry breaking.
The $h_d^0$ eigenstate is then the lowest one and is not much heavier than
$m_Z$, and the residual scalar components of $H_d$ get eaten by the $W$ and
$Z$. The physical charged Higgs particle and neutral pseudoscalar are mostly
components of $H_u^0$ and have mass close to $h_u^0$.  

The $H_u^0$ is then a physical doublet which participates very little
in electroweak symmetry breaking.
Since it is strongly coupled to the top quark ($x=\sin\beta\ll 1$)
it could generate predictions for $b\to s\gamma$ and $R_b$
which are in disagreement with experimental measurements.  Therefore, the Higgs
mass must be sizeable.  To see how large the charged Higgs
mass must be to avoid these constraints, we have plotted in 
Figs.~\ref{bsg} and~\ref{rbhiggs} the effects due to small $x=\sin\beta$.
%%%%%%%%%%%%%%%%%%%%%%%%%%%%%%%%%%%%%%%%%%%%%%%%%%%%%%%%%%%%%%%%%%%%%%
\jfig{bsg}{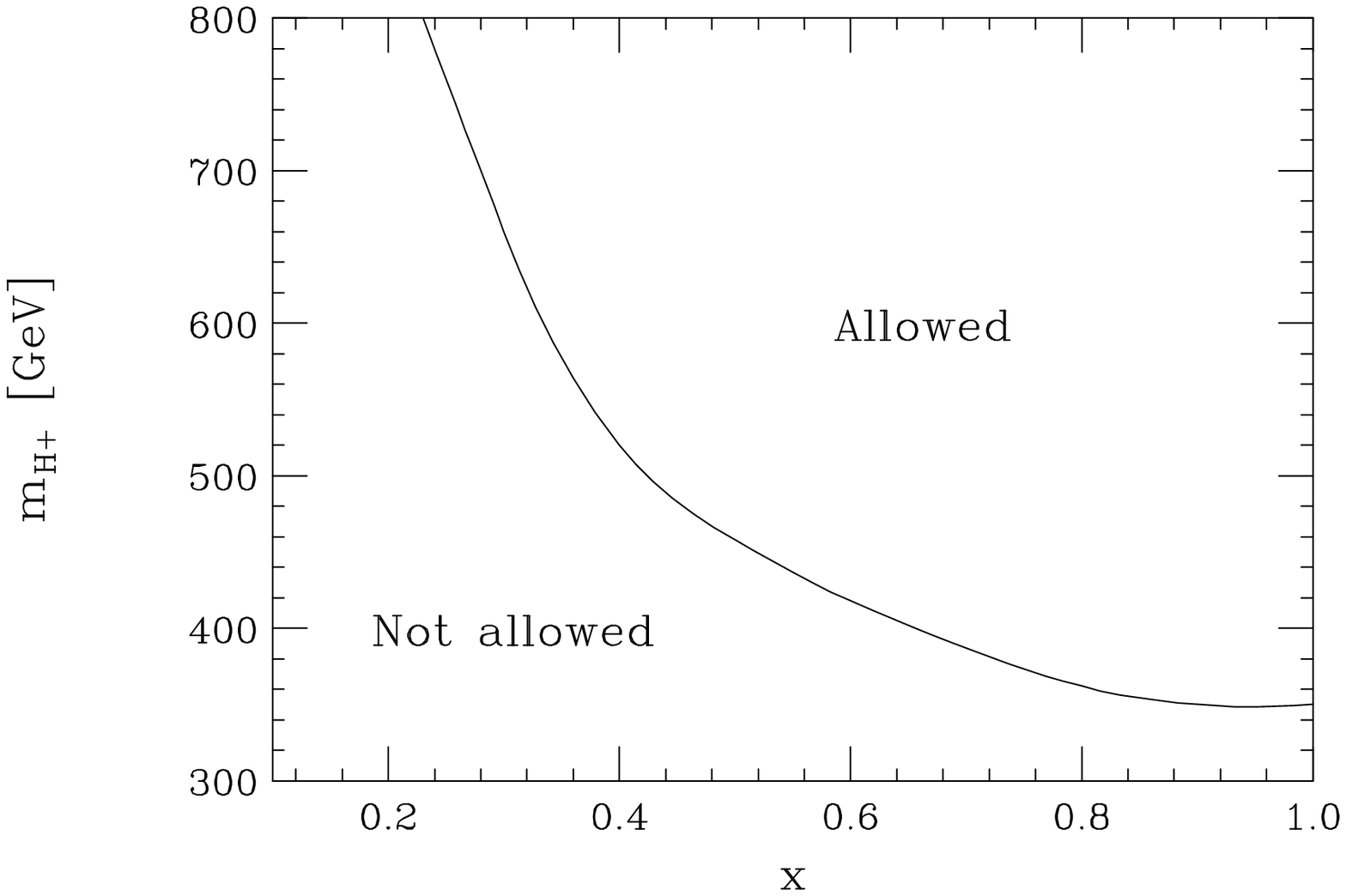}{Limits on the charged Higgs mass versus $x=\sin\beta$
obtained~\cite{r37} from comparing
the measured $B(B\to X_s\gamma)$ rate to the predicted rate.}
%%%%%
\jfig{rbhiggs}{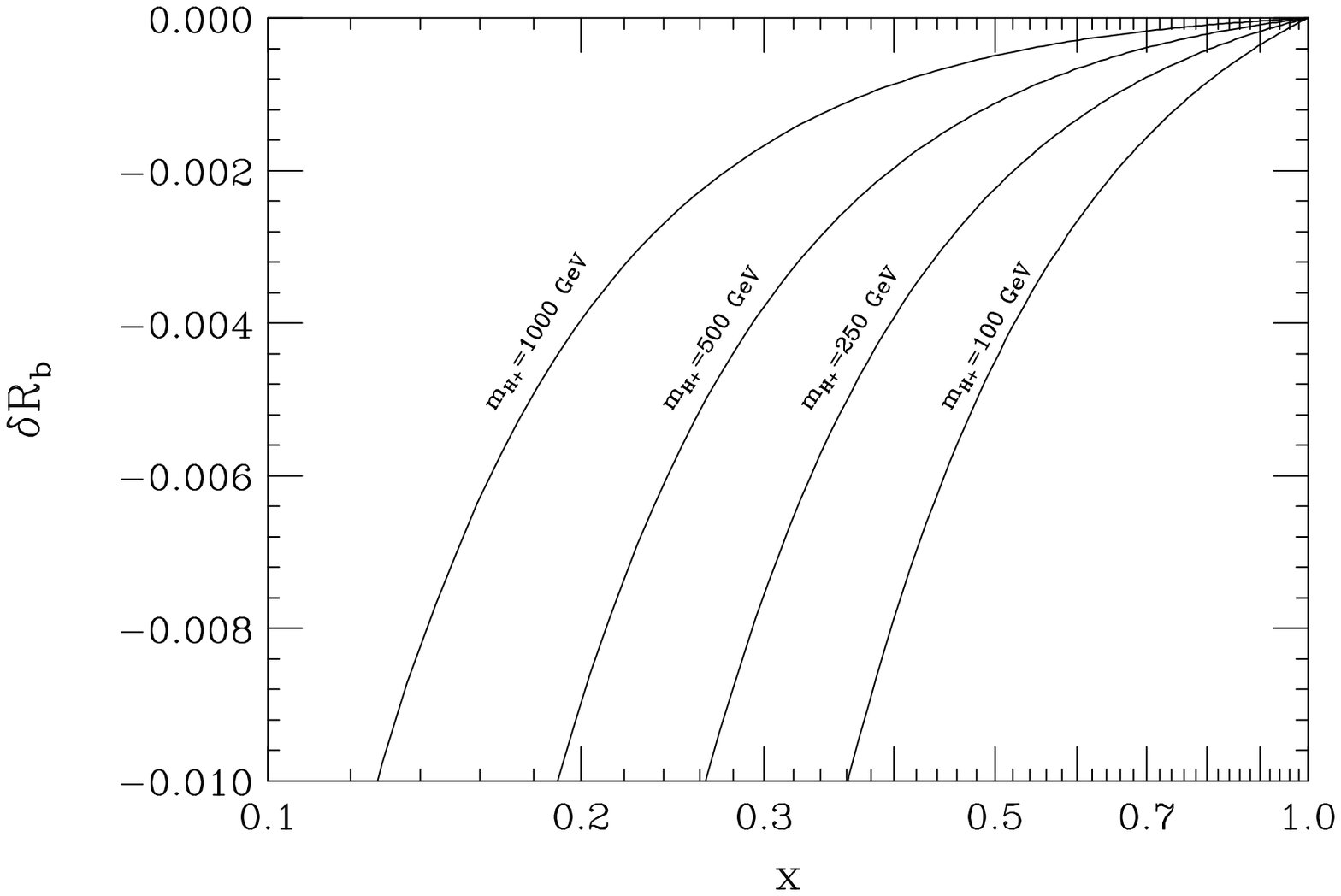}{Contours of $\delta R_b$ versus $x=\sin\beta$
for four values of $m_{H^+}$.  Comparing experimental data to
the theoretical prediction requires that $|\delta R_b|\lsim 0.002$.}
%%%%%%%%%%%%%%%%%%%%%%%%%%%%%%%%%%%%%%%%%%%%%%%%%%%%%%%%%%%%%%%%%%%%%%%%
In Fig.~\ref{bsg} we have indicated the lower limit on $m_{H^+}$ versus
$x$ in order to be consistent with the currently measured 
$B\to X_s\gamma$ rate~\cite{r37}.  In Fig.~\ref{rbhiggs} we plot the
negative shift in $R_b=B(Z\to b\bar b)/B(Z\to {\rm had})$ 
due to the charged Higgs vertex corrections.  The current 
measurement~\cite{r38}
is $R_b=0.2170\pm 0.0009$ which should be compared with the SM prediction
of $R_b=0.2156\pm 0.0003$. Therefore, any $|\delta R_b|\gsim 0.002$
is probably ruled out by the data.  We should also note that in
both $B(B\to X_s\gamma)$ and $R_b$ supersymmetric diagrams
involving $\tilde t_{1,2}$ and $\tilde \chi^\pm_{1,2}$ can also contribute
with the opposite sign and cancel the charged Higgs 
contributions~\cite{r39,r40}.
This would allow smaller charged Higgs masses, and therefore smaller
$m_{h^0_u}$ than Figs.~\ref{bsg} and~\ref{rbhiggs} seem to allow.

Thus we conclude that the generic prediction in supersymmetric models is
that if $h_u^0$ is strongly interacting with the top quark, it must
be accompanied by three scalar particles with similar mass, and that this whole
multiplet must be quite heavy in order to not disrupt $b$-quark observables
too drastically.  This heavy mass prediction 
implies that $h^0_u$ could decay
into numerous supersymmetric particles and also to the lighter
$h^0_d$, thereby complicating the analysis.  For a strongly 
coupled supersymmetric
Higgs boson with mass above $2m_t$, however, these additional decays
are all expected to be weakly coupled in this limit 
except the top quark mode and possibly the top squark mode.  
Also, since the top squarks
receive mass from supersymmetry breaking their masses could be
larger than all the Higgs masses making it kinematically impossible
for the Higgs particle to decay into them.  This is not the most likely
possibility since renormalization group effects tend to push
the top squarks to small masses when the top Yukawa coupling is large.
The $h_u^0$ interacts
with the top squarks via the $A$-term interactions:
$\lambda_t A_t h^0_u\tilde t_L\tilde t_R$. However, the $A$ terms can be
greatly suppressed due to an approximate $R$ symmetry, which is
often present in gauge mediated models, for example~\cite{r41}.    
Therefore, it is likely that if a Higgs boson in the MSSM were strongly
coupled to the top quark it would be quite heavy and its decays
would be dominated by the top quark modes. In general
Higgs decays into top squarks and the myriad subsequent cascade decays
are also possible, however we will not consider it further here.
In the $h_u^0$ eigenstate limit discussed above, the behavior of this
Higgs boson is similar to that of the bound state scalar in top condensation.
We next turn our attention to this model with the realization that
supersymmetric models can yield a very similar phenomenology.

%%%%%%%%%%%%%%%%%%%%%%%%%%%%%%%%%%%%%%%%%%%%%%%%%%%%%%%%%%%%%%%%%%%%
\section{Top-quark condensation}

Our main example is that of low scale top-quark 
condensation~\citer{r2,r49}.
These models allow more varied predictions for the top quark coupling
to the Higgs boson and more naturally yield top quark mass generation
while only mildly affecting electroweak symmetry breaking.
Of course, spontaneously generating a chiral fermion mass must
necessarily spontaneously
break electroweak symmetry.  However this gauge symmetry
breaking may be weak, just as in the case of the light quark condensate
of chiral symmetry breaking in QCD.  If the decay constant associated
with the pions of top-quark condensation is small compared to
the requirements of electroweak symmetry breaking ($f_{\pi_t}\ll v
\approx 174\gev$)
then the top-quark can still get its full mass and yet be strongly
coupled to the condensing bound-state scalar.  We call this condensing
bound-state scalar the ``top Higgs boson''.

The mass of the top Higgs boson is expected to be near
$m_{\htz}=2m_t$~\cite{r4,r13,r17,r42}.  
The top Higgs boson is one state in an $SU(2)$ doublet,
and thus is accompanied by scalar fields
($\pi_t^0$, $\pi_t^\pm$).  Depending on the decay constant they may
be eaten by the $W^\pm$ and $Z$ vector bosons ($f_{\pi_t}=v$) or
are physical eigenstates ($f_{\pi_t}\ll v$).  In this latter
case the electroweak 
symmetry must be broken by some other mechanism (e.g.,
ETC interactions or
a fundamental scalar) and the top pions become 
pseudo-goldstone bosons whose mass depends on the amount of the
explicit chiral breaking and also on the scale of top-quark 
condensation~\cite{r10,r13}.
For this reason, we will not consider the effects from the top pions.
It should be noted, however, that
if the top pions are light they in general could mediate dangerous
flavor changing neutral currents~\cite{r13,r12}
or too small $R_b$~\cite{r12a},
just as the charged Higgs particle does in
supersymmetric models with low $\tan \beta$.
We assume that a combination of the pions being 
heavy\footnote{Small corrections to the NJL approximations and
a larger ``explicit'' top quark mass (e.g., from ETC interactions)
can substantially increase $m^2_{\pi_t}$.} and aligning
according to some flavor symmetries solves these problems.  For our
purposes including their effects would only increase the signal
event rates that we discuss in the next section. 

The $\pi_t$ decay constant
and the dynamical top quark mass can be related to each other
by the Pagels-Stokar formula~\cite{r10}:
\beq
\label{pagels}
{f^2_{\pi_t}}\simeq \frac{N_c}{16\pi^2}m_t^2\log 
        \frac{\Lambda^2}{m_t^2}
\eeq
where $\Lambda$ is the top-quark condensate scale.
Finetuning considerations in the gap equation for the top-quark mass
imply that $\Lambda$ should probably not be much larger than about 
1 TeV~\cite{r10,r11,r47}.
For this value of $\Lambda$ one expects $f_{\pi_t}/v\simeq 1/4$.

It is not clear how much one should trust the quantitative results
of the gap equation, the Pagels-Stokar formula, or any other equation
which attempts to be precise in the strong coupling regime.  
For this reason, it is perhaps best to treat $x\equiv f_{\pi_t}/v$ 
and $m_{\htz}$ as free parameters.  Unless otherwise indicated, however,
we will use $x=1/4$ in sympathy with the standard approximation
schemes and fine-tuning considerations.

%%%%%%%%%%%%%%%%%%%%%%%%%%%%%%%%%%%%%%%%%%%%%%%%%%%%%%%%%%%%%%%%%%%
\section{Collider signatures}

The Tevatron
and LEPII will both have difficulties detecting $\htz$
because both
of these colliders rely on ``gauge coupled'' production modes
such as $e^+e^-\to Z\htz$ and $q\bar q'\to W\htz$.  Since $\htz$ couples
weakly to gauge bosons, it is not likely that the Tevatron would see it
even if it were light.\footnote{Since the $gg\to \htz$ process is enhanced by
$1/x^2$, the dominant gluon
fusion mechanism may become visible at the Tevatron. This, however, requires
a detailed detector simulation.}
With enough luminosity it is possible that LEPII could see $\htz$
in its ordinary Higgs searches.
The cross section for $\htz$ is
\beq
\sigma(e^+e^-\to Z\htz)=x^2\sigma(e^+e^-\to Z\hsmz).
\eeq
Therefore, if $x$ is low LEPII might miss this state
even if it is kinematically accessible.  Unlike LEPII and Tevatron,
the prospects for $\htz$ discovery at LHC increase as $x$ decreases.
For this reason we focus on LHC observables.

The top Higgs boson couples appreciably only to $t\bar t$ and somewhat more
weakly to $WW$ and
$ZZ$ pairs.  However, the coupling to the top quark is enhanced
by a factor of $1/x$ in comparison with the top quark coupling to the
Standard Model Higgs boson~\cite{r10,r15}.  This is obvious since
\beq
\lambda_t\simeq \frac{m_t}{f_{\pi_t}} = \frac{1}{x}\frac{m_t}{v}
 = \frac{\lambda_t^{SM}}{x}.
\eeq 
Similarly, the coupling to the vector bosons is suppressed by a factor of $x$.
The branching fractions are presented in Fig.~\ref{br}.
%%%%%%%%%%%%%%%%%%%%%%%%%%%%%%%%%%%%%%%%%%%%%%%%%%%%%%%%%%%%%
\rfig{br}{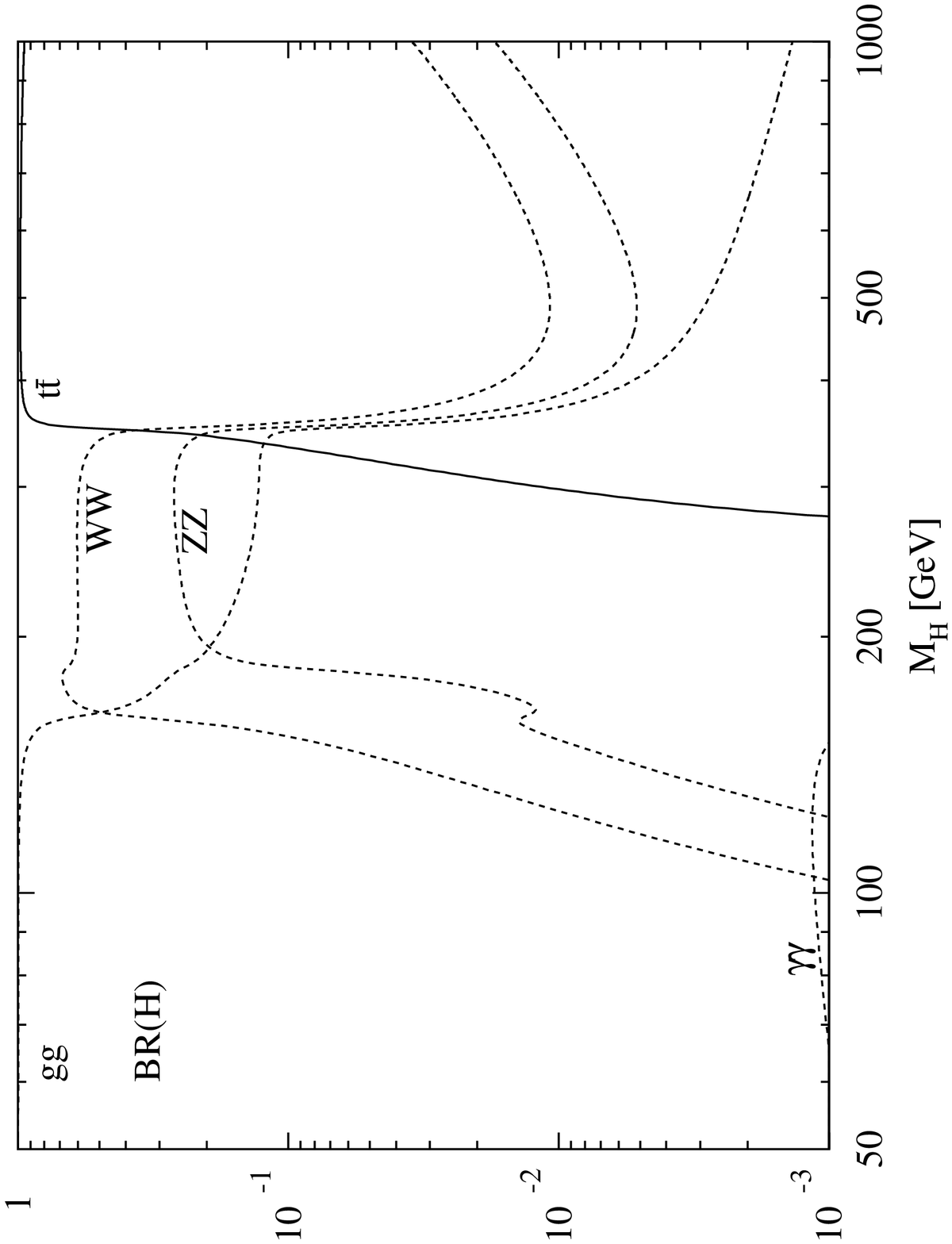}{Branching fractions of $H\equiv {\htz}$ versus its mass
for $x\equiv f_{\pi_t}/v=1/4$.}
%%%%%%%%%%%%%%%%%%%%%%%%%%%%%%%%%%%%%%%%%%%%%%%%%%%%%%%%%%%%%
All relevant radiative corrections are included according to
Ref.~\cite{r44a,hdecay}.

One striking difference between the top Higgs boson and the
SM Higgs boson is the lack of a $b\bar b$ channel. Thus, $b$-tagged
final states are not important for a light $h_t^0$.  Furthermore, when
$m_{\htz}>2m_t$ the branching ratio into $t\bar t$ is 
close to 100\%, which should be compared with the SM Higgs boson, whose
$WW$ decay mode is always greater than $t\bar t$ for any mass~\cite{r44a,r43}.
The large $t\bar t$ branching fraction for $\htz$ arises since
its partial width is enhanced by $1/x^2$ and the $WW$ partial width 
simultaneously decreases by $x^2$.

The production cross sections are also quite unusual compared
to the Standard Model. In Fig.~\ref{lhc} we plot the production
cross sections of $\htz$ in different modes for the LHC with
$\sqrt{s}=14\tev$.
%%%%%%%%%%%%%%%%%%%%%%%%%%%%%%%%%%%%%%%%%%%%%%%%%%%%%%%%%%%%%%%%%%%%%
\rfig{lhc}{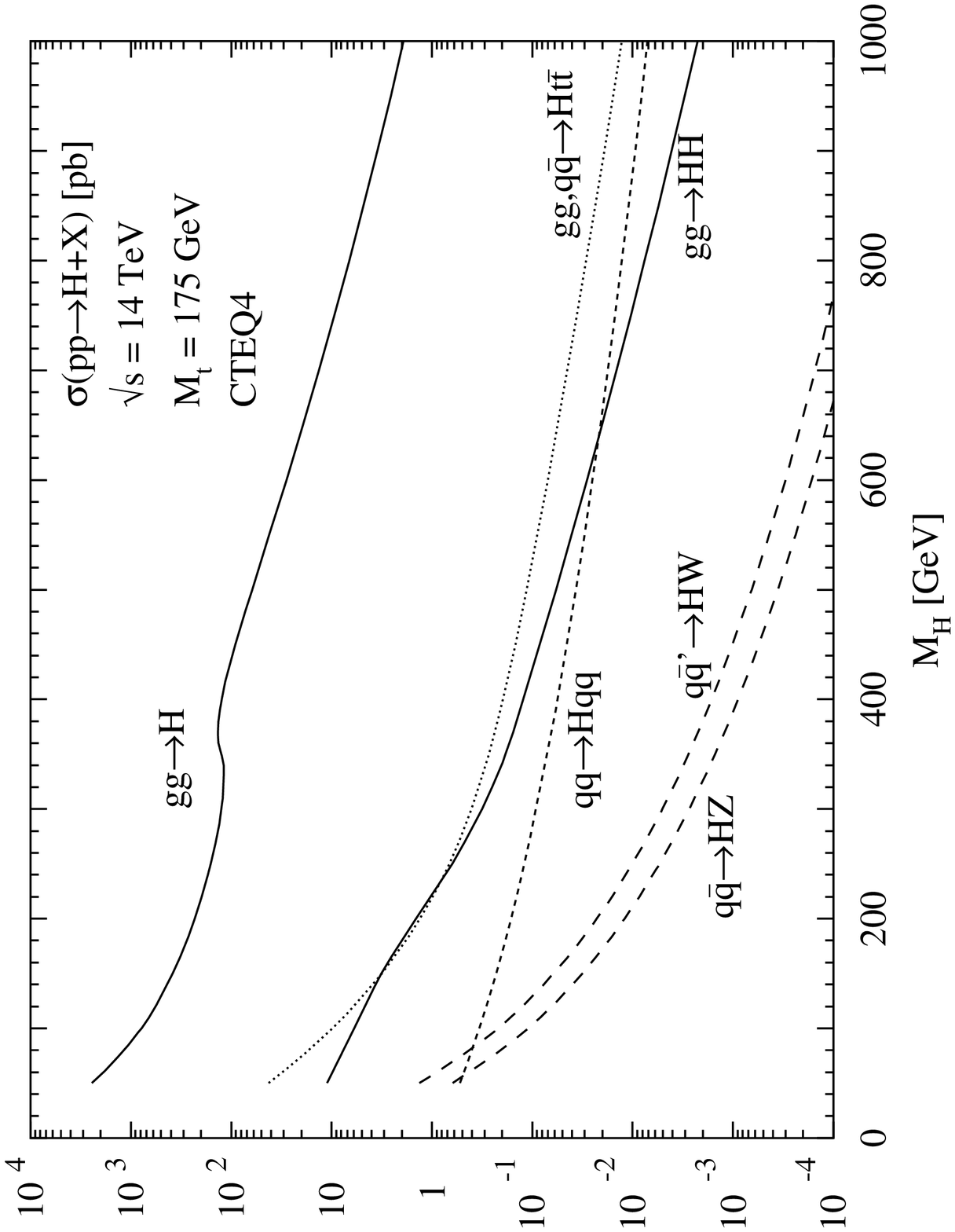}{Production cross sections of the top Higgs boson $H\equiv
\htz$ at the LHC with with $\sqrt{s}=14 \tev$ for $x\equiv f_{\pi_t}/v=1/4$.}
%%%%%%%%%%%%%%%%%%%%%%%%%%%%%%%%%%%%%%%%%%%%%%%%%%%%%%%%%%%%%%%%%%
We include the full NLO QCD corrections to $gg\to \htz$ \cite{r44a,glufus},
$qq\to \htz qq$ \cite{vvh} and $q\bar q\to \htz V~[V=W,Z]$ \cite{vhv}.
In these cross sections we adopted the CTEQ4M parton densities \cite{cteq4}
with $\alpha_s^{NLO}(M_Z^2) = 0.116$. The QCD corrections to the processes
$gg,q\bar q \to \htz t\bar t$ and $gg\to \htz \htz$ are unknown, so that we
evaluated them using CTEQ4L parton densities with $\alpha_s^{LO}(M_Z^2)=0.132$.
The ``gauge processes'' such as
$q\bar q\to \htz Z$ and $q\bar q'\to \htz W^\pm$ are smaller
by $x^2$ in comparison to the Standard Model, and the ``Yukawa
processes'' such as $gg\to \htz$ and $gg,q\bar q\to \htz t\bar t$
are enhanced by $1/x^2$.  Note also that the $gg\to \htz\htz$
production process is enhanced by a factor of $1/x^4$ with respect
to the Standard Model. [For $x=1/4$ one finds a factor of 256 
enhancement above the same process in the Standard Model with equal 
Higgs boson mass.]
In the Standard Model the two Higgs production cross section
is not large enough to play a significant role in the phenomenology
at the LHC~\cite{r44}.  
However, when the Higgs boson strongly interacts
with the top quark this process is most affected and becomes
relevant. We will make use of this later.

At the LHC the ``gold-plated'' Higgs discovery mode for the
SM Higgs boson in the mass range $140\gev \lsim m_{\hsmz}\lsim 800\gev$ is
the $gg\to \hsmz\to ZZ^{(*)}\to 4l^\pm$ signature~\cite{r43}.  Although the
$gg\to \htz$ 
cross section has been enhanced by a factor of $1/x^2$ in the top
Higgs case, it turns out that this mode would be more difficult than it
is for the SM Higgs boson for two reasons: First,
the branching fraction into $ZZ$ decreases by {\it more} than $x^2$
beyond the $t\bar t$ threshold
since the partial width into $ZZ$ decreases by $x^2$ and the total
Higgs width increases because of the enhanced $t\bar t$ partial width.
The total rate of $gg\to \htz\to ZZ$ is plotted in Fig.~\ref{br2}.
%%%%%%%%%%%%%%%%%%%%%%%%%%%%%%%%%%%%%%%%%%%%%%%%%%%%%%%%%%%%%%%
\rfig{br2}{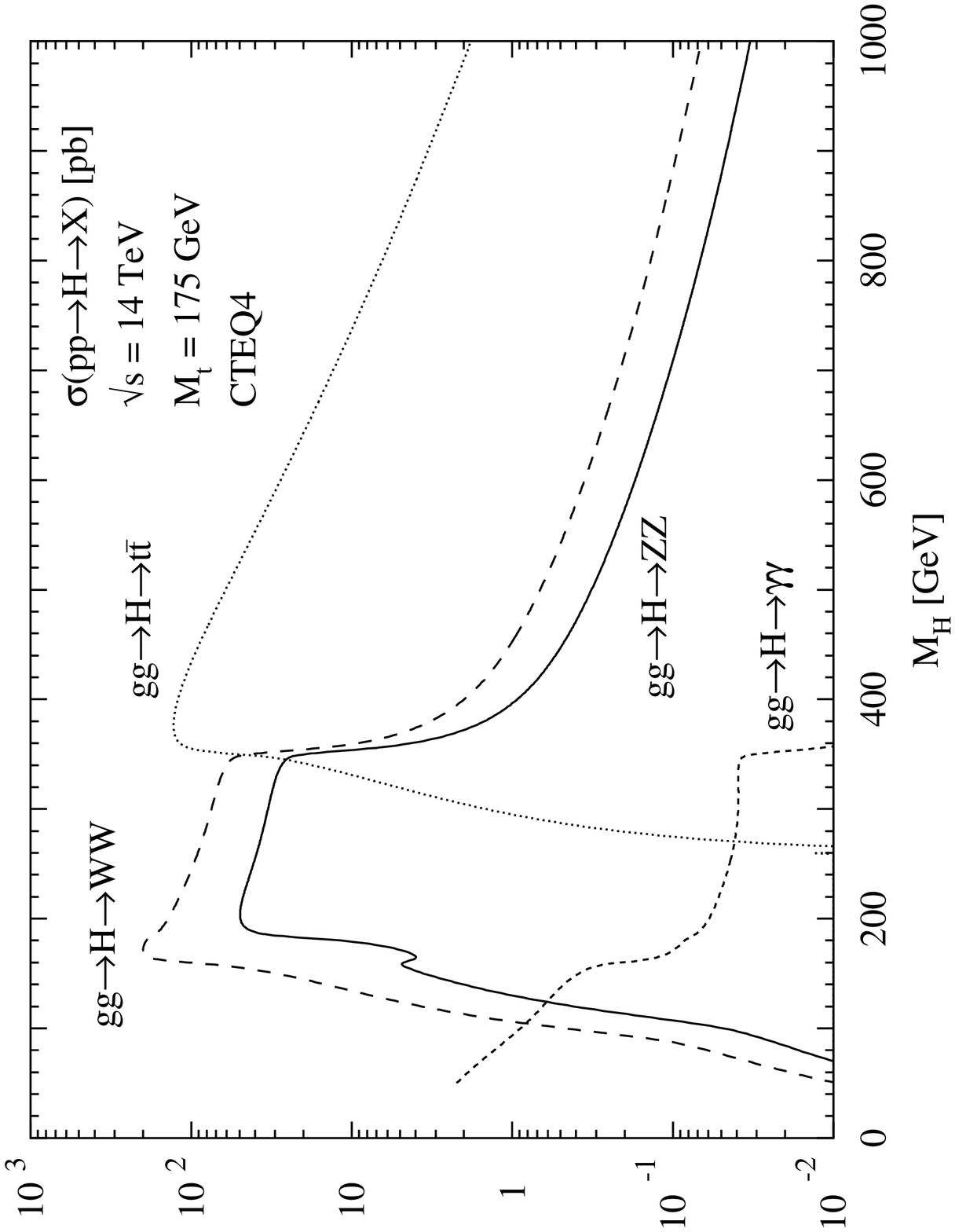}{Cross sections for the production and two-body decays of
the top Higgs boson $H\equiv \htz$ for $x\equiv f_{\pi_t}/v=1/4$.}
%%%%%%%%%%%%%%%%%%%%%%%%%%%%%%%%%%%%%%%%%%%%%%%%%%%%%%%%%%%%%%
Second, since the total width of the Higgs particle is significantly larger
above the $t\bar t$ threshold
due to the increased coupling of the top Higgs boson to the top quarks, the
invariant mass ``bump'' of the $h\to ZZ\to 4l^\pm$ is no longer narrow, but
quite broad.  The Higgs width is already at about $70\gev$ for
$m_{\htz}=400\gev$ (see Fig.~\ref{gam}), while the
%%%%%%%%%%%%%%%%%%%%%%%%%%%%%%%%%%%%%%%%%%%%%%%%%%%%%%%%%%%%%%%%
\rfig{gam}{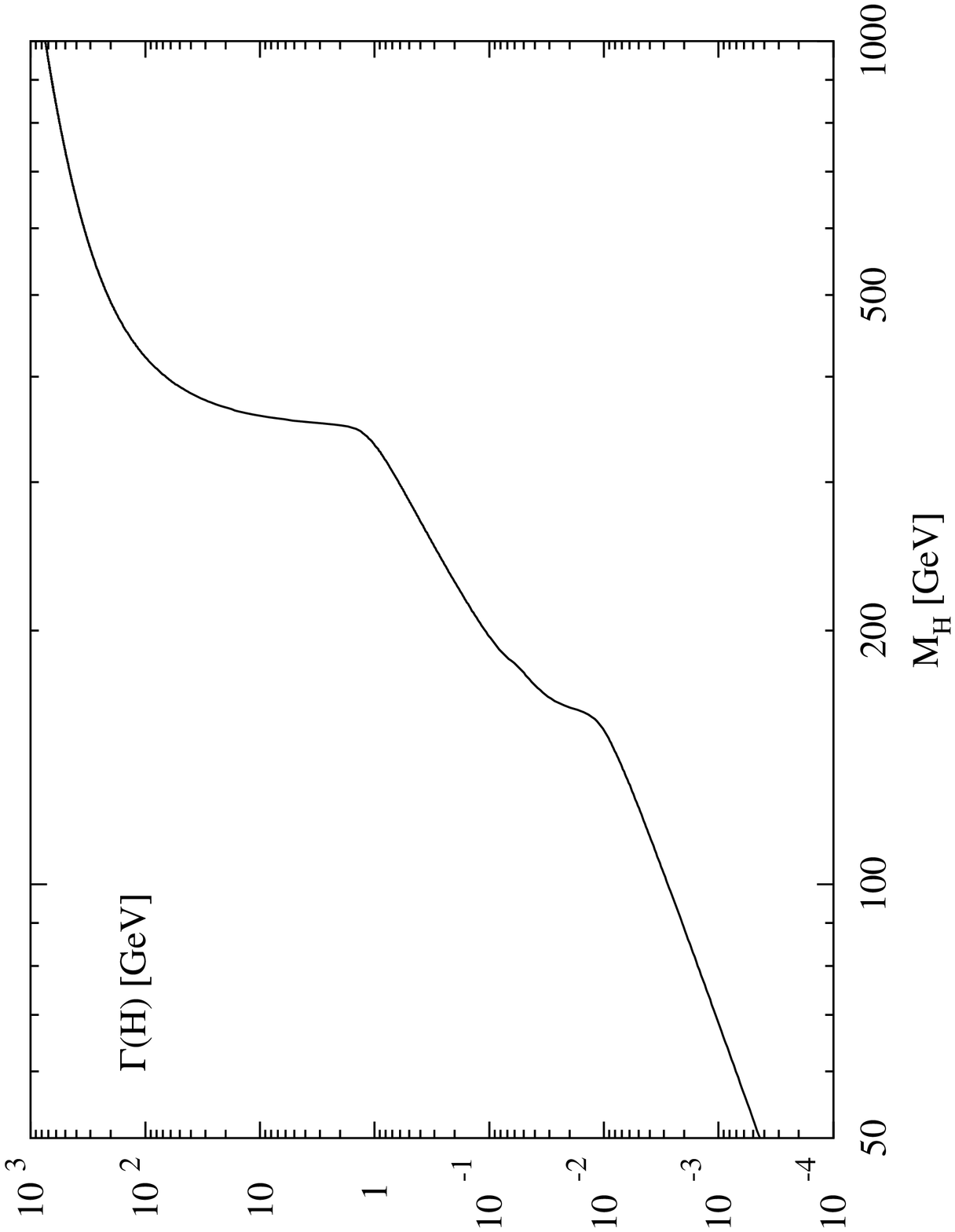}{Total width of $H\equiv \htz$ as a function of its mass
for $x\equiv f_{\pi_t}/v=1/4$.}
%%%%%%%%%%%%%%%%%%%%%%%%%%%%%%%%%%%%%%%%%%%%%%%%%%%%%%%%%%%%%%%%%
SM Higgs particle has a width of less than $30\gev$
for the same mass.  As the mass of the top Higgs scalar gets larger the
problem worsens, and finding a bump above the background
is more difficult. For low values of $x$ the gold-plated mode is probably only
useful for Higgs masses between about $180\gev$ and $400\gev$.

Another useful signature is the $gg\to \htz\to WW$ leptonic decay
mode~\cite{r52,r27} in the 
mass region of $160\gev \lsim m_{\htz}\lsim 340\gev$.
This is also plotted in Fig.~\ref{br2}.
In this range, $\htz\to WW$ is  competing successfully with 
loop mediated decays into $gg$ or off-shell $t^* t$ decays.  Therefore,
the production cross section $gg\to \htz$ is enhanced by the strong top
interactions, but the branching fraction is not significantly suppressed.
In this region the methods of ref.~\cite{r27} can be used to extract
a signal and furthermore perhaps even extract the Higgs mass.

For a light SM Higgs boson ($m_{\hsmz}\lsim 150\gev$) it has been 
found~\cite{r43}
that the most useful signature for the Higgs search is $gg\to
\hsmz\to\gamma\gamma$.
The cross section varies between $35\xfb$ and $85\xfb$ 
in the mass range $100\gev < m_{\hsmz}< 150\gev$.  For the top Higgs boson,
the cross section is at about $1\xpb$ for $m_{\htz}=100\gev$ and stays
above $40\xfb$ for $m_{\htz}< 340\gev$ as can be seen
in Fig.~\ref{br2}. Therefore, photon final states
can be a very useful signature of the top Higgs particle at mass scales twice
as large as the applicable region for the SM Higgs boson.  This result depends
crucially on the fact that $\htz$ is a pure mass eigenstate and does not
mix with other electroweak symmetry breaking mechanisms, especially those
that give mass to the bottom quarks.  Even with some mixing of other
sectors with the top Higgs boson the photon final states should still
be enhanced over that of the Standard Model.  Also, in the topcolor
models of top-quark condensation, the instanton associated with
the strongly coupled topcolor gauge group could generate part or
all of the bottom quark mass~\cite{r10,r13}.  This also will 
mediate an effective coupling between the top Higgs particle and the bottom
quarks which could even be larger than the Standard Model Higgs coupling
to bottom quarks.  Clearly, such a large coupling to the bottom quarks
if present would spoil
the $\gamma\gamma$ signal for the top Higgs boson.

The Higgs mediated production of $t\bar t$ is enhanced by more than
a factor of $1/x^2$ for the top Higgs scalar. It might be possible to utilize
this excess to extract a signal against the background~\cite{r48}.  The maximum
expected Higgs mediated cross section of $t\bar t$ production is
$100\xpb$ at a top Higgs mass just below $400\gev$.  Although this
is a large cross section it is still more than an order of magnitude
below the enormous Standard Model background of about $2\, {\rm nb}$
for $m_t=175\gev$.  Since we can always find other modes more significant
than $t\bar t$ we do not consider it further here.

Perhaps the most striking signal for the top Higgs boson is $4W+X$ production.
All the production modes leading to this final state are shown
in Fig.~\ref{br4}.
%%%%%%%%%%%%%%%%%%%%%%%%%%%%%%%%%%%%%%%%%%%%%%%%%%%%%%%%%%%%%%%%%%%%%
\rfig{br4}{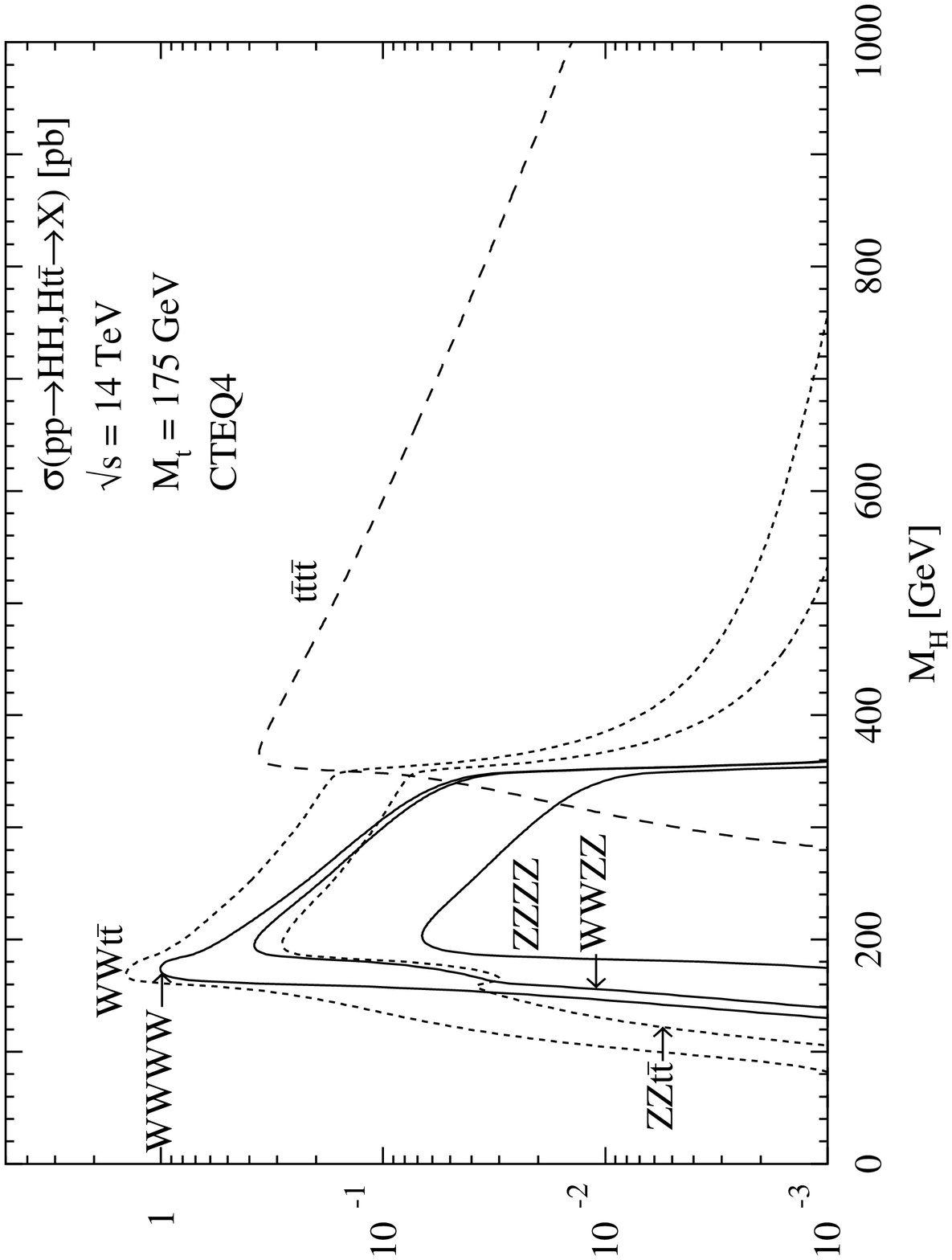}{Cross sections for the production and four-body decays of
the top Higgs boson $H\equiv \htz$ for $x\equiv f_{\pi_t}/v=1/4$.}
%%%%%%%%%%%%%%%%%%%%%%%%%%%%%%%%%%%%%%%%%%%%%%%%%%%%%%%%%%%%%%%%%%%%
The underlying processes contributing to these modes include
$pp\to \htz t\bar t\to t\bar t t\bar t, WWt\bar t$
and $pp\to \htz\htz\to t\bar tt\bar t, WWt\bar t, WWWW$
and the top quarks subsequently decay into $Wb$.  In addition to
these modes there are four gauge boson modes involving the $Z$ bosons.

The Standard Model background for these processes is small.  The largest
source of $4W+X$ in the Standard Model is $gg\to t\bar t t\bar t$
which has a cross section less than $10\xfb$~\cite{r32,r1}.  
If $x$ is near one then the top Higgs boson acts very similar to the
Standard Model Higgs particle and contributes less than $2\xfb$ to the
four top rate.  The maximum rate occurs if $m_{\hsmz}\simeq 450\gev$.
However, the top Higgs boson with smaller $x$
can exceed the Standard Model rate by more than
one order of magnitude.  With $x=1/4$, as is plotted in
Fig.~\ref{br4}, even a $1\tev$ Higgs mass contributes more than
$10\xfb$ to the four top rate.
And when
the Higgs mass is at its preferred value of $m_{\htz}=2m_t$ we
find a four top cross section in excess of $100\xfb$.  
We note also that in topcolor there are other sources of four top
production.  Namely, if the gauge bosons associated with the topcolor
gauge group are light enough they can be pair produced and will subsequently
decay into four top quarks~\cite{r45}. 

Reconstructing the 
mass is a bit ill-defined task for the heavier top Higgs boson
since its width is greater than $100\gev$
when its mass is above $\sim 400\gev$ (see Fig.~\ref{gam}).  If the mass is
at its preferred value of $m_{\htz}=2m_t$ or lighter then 
the width is still narrow enough (less than $5\gev$) such
that extracting a top Higgs mass is possible in principle.  

If the top Higgs mass is in the range of 
$170\gev\lsim m_{\htz}\lsim 340\gev$ the $4W$ and $WWt\bar t$
modes dominate.  The total expected cross section of
four vector bosons in the Standard Model from $WWWW$ and $WWt\bar t$
modes is less than $20\xfb$~\cite{r32}. The SM rate from just $WWWW$
is about $1\xfb$~\cite{r32}.  Therefore, the large enhancement of 
$WWWW$ production with zero or two $b$-jets is a good signal for an
intermediate top Higgs boson.  Mass reconstruction might be possible
in this region using similar techniques as in Ref.~\cite{r27}; however,
combinatoric
ambiguities of the final state leptons with a $4W$ final state may subvert
such attempts.  Furthermore,  the two different production topologies
of $t\bar t\htz$ and $\htz\htz$ each contribute to the $WWt\bar t$ final
state signal.  This further complicates the analysis for the $4W$ plus
two $b$-jet mode, and more detailed Monte Carlo simulations are
required to find out the maximal amount of information that can
be extracted.

%%%%%%%%%%%%%%%%%%%%%%%%%%%%%%%%%%%%%%%%%%%%%%%%%%%%%%%%%%%%%%%%%%%%%%
\section{Conclusion}

Table 1 is a summary of the most useful modes to discover the top Higgs boson
for various mass regions.  
%%%%%%%%%%%%%%%%%%%%%%%%%%%%%%%%%%%%%%%%%%%%%%%%%%%%%%%%%
\begin{table}
\begin{center}
\begin{tabular}{ccc}
\hline \hline 
mode & mass range & 
{\begin{tabular}{c}
Is mass \\
reconstructable? 
\end{tabular}} \\ 
\hline 
$gg\to\htz\to\gamma\gamma$ & $m_{\htz}\lsim 340\gev$ & $\surd$ \\ 
\vspace{0.05in}
$gg\to\htz\to WW^{(*)}$ & $150\gev\lsim m_{\htz}\lsim 340\gev$ & 
       $\surd$ \\ 
\vspace{0.05in}
$gg\to\htz\to ZZ$ & $180\gev\lsim m_{\htz}\lsim 400\gev$ & $\surd$ \\
\vspace{0.05in}
$gg\to\htz\htz\to WWWW$ & $150\gev\lsim m_{\htz}\lsim 340\gev$ & ? \\
\vspace{0.05in}
{\begin{tabular}{c}
$gg,q\bar q\to \htz t\bar t\to WWt\bar t$ \\
$gg\to\htz\htz\to WWt\bar t$
\end{tabular}}\Huge\} &  $150\gev\lsim m_{\htz}\lsim 350\gev$ & ? \\
\vspace{0.05in}
{\begin{tabular}{c}
$gg,q\bar q\to \htz t\bar t\to t\bar tt\bar t$ \\
$gg\to \htz\htz\to t\bar tt\bar t$ 
\end{tabular}}\Huge\} & $350\gev\lsim m_{\htz}\lsim 1\tev$ & X \\
\hline\hline
\end{tabular}
\caption{Summary of top Higgs production and decay modes for
$x=1/4$.}
\end{center}
%\label{spectrumtable}
\end{table}
%%%%%%%%%%%%%%%%%%%%%%%%%%%%%%%%%%%%%%%%%%%%%%%%%%%%%%%%%%%%%%%%%%
In the first column we list final states arising through $\htz$ production
and in the second column we list the top Higgs boson mass range for
which this signature is applicable.  
We stress that the large enhancement of the four top rate is a
somewhat unique and spectacular signature of the heavy top Higgs boson.
In fact, the enhancement of the four top mode may be the {\it only}
discernible signal of the top Higgs boson at the LHC.

Many results that we presented in the previous section were based upon
the choice $x=1/4$.  The rates for different values of $x$ can
in principle be extracted from the plots we have provided.  
The production cross sections are straightforward
to generalize for different values of $x$.  We have already noted above
that the $gg\to \htz\htz$ production cross section is enhanced over
the Standard Model by a factor of $1/x^4$.  Similarly, the $gg\to\htz$
cross section is enhanced by $1/x^2$.  The decay branching fractions
are somewhat more complicated.  Some partial widths are suppressed by
$x^2$ and some are enhanced by $1/x^2$.  In Fig.~\ref{s2xt} we plot the cross
sections for the
production and two-body decays and in Fig.~\ref{s4xt} for the production and
four-body decays of $\htz$ as a function of $x$ with
$m_{\htz}=2m_t$ -- the preferred top Higgs boson mass of top-quark
condensate models.
%%%%%%%%%%%%%%%%%%%%%%%%%%%%%%%%%%%%%%%%%%%%%%%%%%%%%%%%%%%%
%\rfig{brxt}{brxt.ps}{The Higgs boson decay branching fractions
%as a function of $x$.  The Higgs boson mass is set to $m_{\htz}=2m_t$,
%the preferred value in top-quark condensation.}
%%%%%%%%%%%%%%%%%%%%%%%%%%%%%%%%%%%%%%%%%%%%%%%%%%%%%%%%%%%%
\rfig{s2xt}{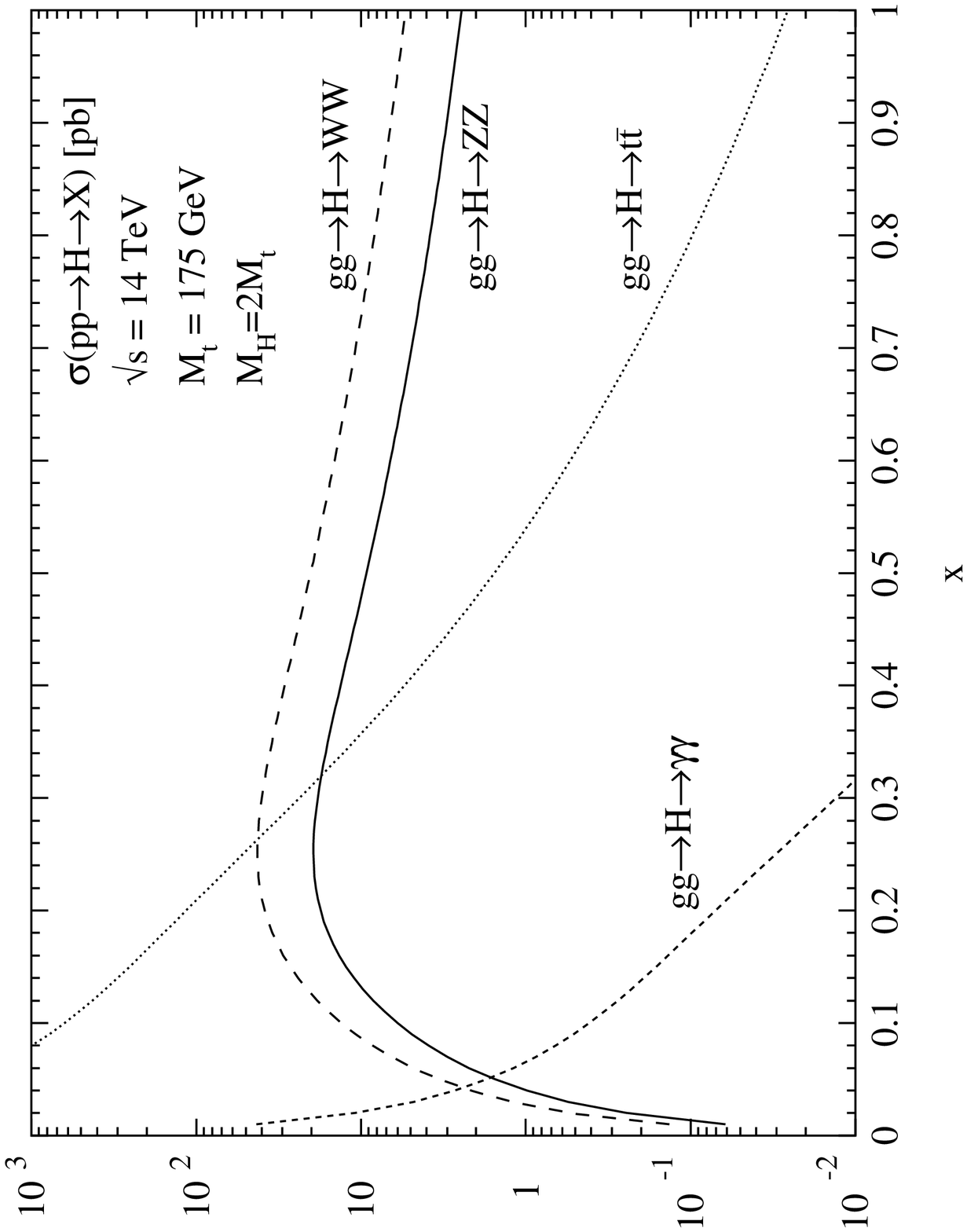}{Cross sections for the production and two-body decays
of the top Higgs boson for $m_{\htz} = 2 m_t$.}
%%%%%%%%%%%%%%%%%%%%%%%%%%%%%%%%%%%%%%%%%%%%%%%%%%%%%%%%%%%%
\rfig{s4xt}{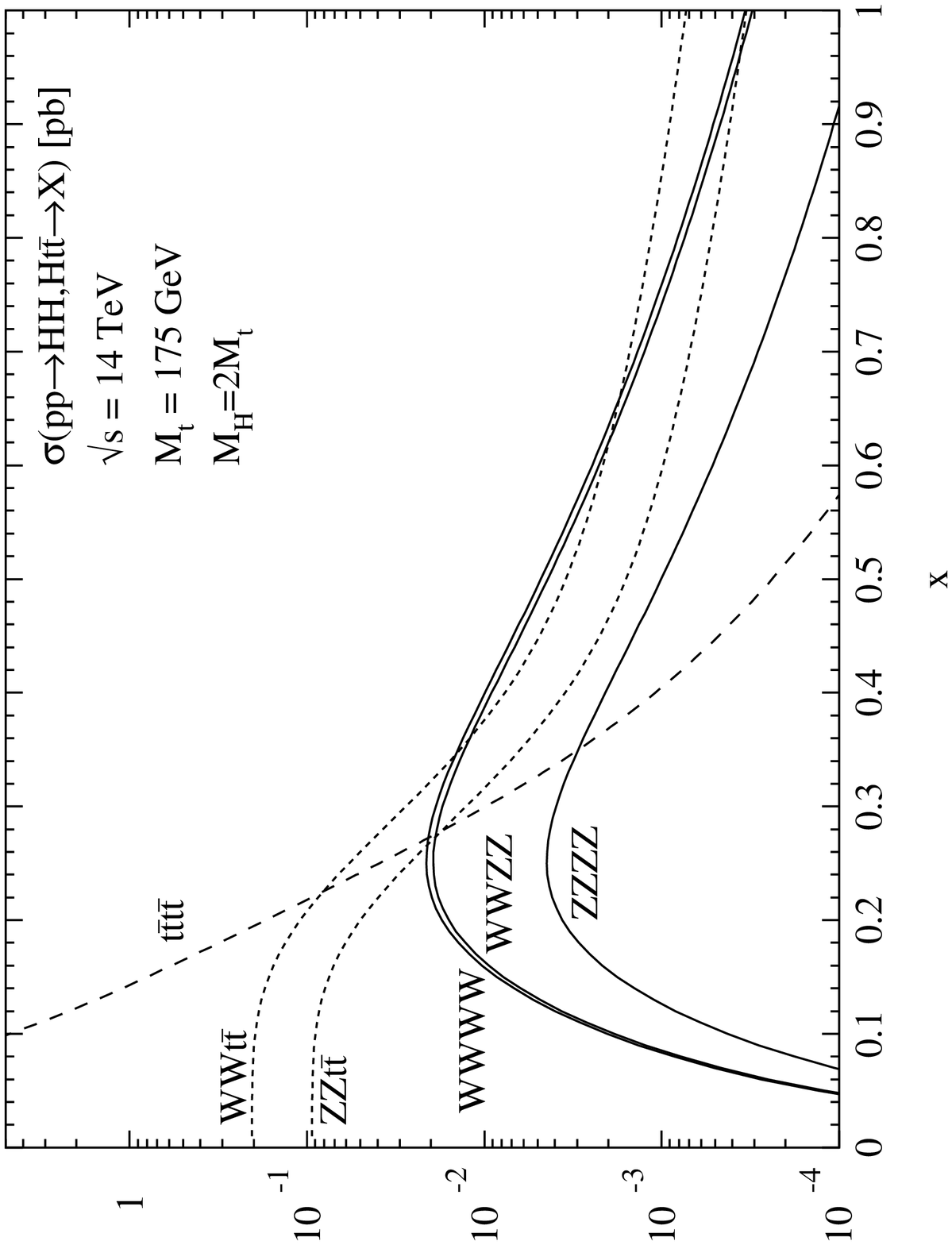}{Cross sections for the production and four-body decays
of the top Higgs boson for $m_{\htz} = 2 m_t$.}
%%%%%%%%%%%%%%%%%%%%%%%%%%%%%%%%%%%%%%%%%%%%%%%%%%%%%%%%%%%%
For $x=1/4$ the $WW$ and $t\bar t$ modes
have almost identical branching fractions.  As $x\to 1$ the branching
fractions tends more and more to the Standard Model values. Thus for the
two-body final states at
small values of $x$ the $t\bar t$ final state dominates, while beyond
$x\sim 0.3$ it is more and more suppressed thus leaving only the $WW,ZZ$ final
states
as visible signals similar to the SM. The $\gamma\gamma$ final state is only
significant for $x\lsim 0.3$.
For larger values of $x$
the $t\bar t \htz$ production mode becomes more important
for the phenomenology of the top Higgs boson at the LHC compared to
$\htz\htz$ production, and the four-body $t\bar tt\bar t$ signal loses
significance.  However, the $\htz\to ZZ$ mode becomes
more important as $x$ goes to 1, and the phenomenology approaches
that of the SM Higgs boson.

Much of the known phenomenology associated with Higgs boson signatures
at high energy colliders has been closely related to the Standard Model.
Even light Higgs collider studies in supersymmetric models often deviate only
mildly from a Standard Model Higgs boson~\cite{r44a,glufus,r46}.  
This is especially true in the MSSM.  However, if electroweak
symmetry breaking and fermion mass generation arise through a more
sophisticated mechanism than in the SM or MSSM,
the relevant Higgs states
may be more difficult to detect, and may require signatures that are
not useful in SM Higgs searches.  For example, in top condensate
models, electroweak symmetry breaking may be accomplished by one Higgs
scalar $\hewz$ and top quark mass generation by another $\htz$. 
In this case, $\hewz$ will be difficult to find~\cite{r47} 
at the LHC and might be
seen only by extracting a three lepton signal above 
background~\cite{r47a}, 
and we have shown above that the top Higgs boson signature might only be
seen through four vector bosons or four top quarks.  This is just one
clear illustration of how a strongly coupled Higgs boson in the spectrum
can dramatically change Higgs phenomenology at high energy colliders.

%%%%%%%%%%%%%%%%%%%%%%%%%%%%%%%%%%%%%%%%%%%%%%%%%%%%%%%%%%%%%%%%%%%%%%%%

\end{document}